\documentclass[preprint,12pt,authoryear]{elsarticle}

\usepackage{amssymb}
\usepackage{longtable}
\usepackage{color}
\usepackage{lscape}
\usepackage{hyperref}

\usepackage{tikz}
\usepackage{tikz-qtree}
\usetikzlibrary{trees}

\journal{}

\begin{document}

\begin{frontmatter}


\cortext[cor1]{Corresponding author}

\title{Collaborative vehicle routing: a survey}


\author[aut1]{Margaretha Gansterer\corref{cor1}}
\ead{margaretha.gansterer@univie.ac.at}
\author[aut1]{Richard F. Hartl}
\ead{richard.hartl@univie.ac.at}

\address[aut1]{University of Vienna, Department of Business Administration, Oskar-Morgenstern-Platz 1, 1090 Vienna, Austria}

\begin{abstract}
In horizontal collaborations, carriers form coalitions in order to perform
parts of their logistics operations jointly. By exchanging transportation requests
among each other, they can operate more efficiently and in a more sustainable
way. Collaborative vehicle routing has been extensively discussed in the literature. We identify three major streams of research: (i) centralized collaborative planning, (ii) decentralized planning without auctions, and (ii) auction-based decentralized planning. For each of them we give a structured overview on the state of knowledge and discuss future research directions.   
\end{abstract}

\begin{keyword}
Logistics \sep Collaborations \sep Vehicle routing 



\end{keyword}

\end{frontmatter}


\section{Introduction}
\label{intro}

The transportation industry is highly competitive and companies need to aim for
a maximum level of efficiency in order to stay in business. Fierce competition 
brings prices down and therefore profit margins have declined to an extremely low level. To increase efficiency, these companies
can establish collaborations, where parts of their logistics operations are planned jointly. By \textit{collaborative vehicle routing} we refer to all kinds of cooperations, which are intended to increase the efficiency of vehicle fleet operations.\footnote{We use the terms \textit{collaboration} and \textit{cooperation} interchangeable. In the literature, there is an agreement that collaboration is a strong type of cooperation. However, the boundary between them is vague (\citealp{CruijsenSurv}).} By increasing efficiency, collaborations also serve ecological goals. It is well known that transportation is one of the main contributors of $CO_2$ emissions (\citealp{ballot10}). Thus, public authorities are encouraging companies to collaborate. They not only aim at reduced emissions of harmful substances, but also on reduced road congestion, and noise pollution. Moreover, collaborations in logistics have been shown to increase service levels, gain market shares, enhance capacities, and reduce the negative impacts of the bullwhip effect (\citealp{audy12}). Thus, it is not surprising that collaborative vehicle routing is an active research area of high practical importance. 

Related reviews by \cite{verdonck13} and \cite{CruijsenSurv} exist. Both are dealing with transportation collaborations. However, \cite{verdonck13} focus on the operational planning of road transportation \textit{carriers} (i.e. the owners and operators of transportation equipment) only. The perspective of collaborating \textit{shippers} (i.e. the owners of the shipments) is not taken into account. Furthermore, they do not consider studies on centralized planning (i.e. collaboration in case of full information). We observe that about 45\% of the related literature refers to central planning situations. This is an important aspect of collaborative vehicle routing, where a centralized authority is in charge of allocating requests such that requirements of all collaborators are met. Furthermore, we identify two classes of decentralized settings, which are auction-based and non-auction-based collaborations.

\cite{CruijsenSurv} give an overview on different types of horizontal collaboration, i.e. the levels of integration among collaborators. They do not consider operational planning problems. We find that almost 60\% of the related articles were published in the last 3 years. These articles have not been covered in both of the existent reviews. 

The review by \cite{Guajardosurvey} deals with cost allocation in collaborative transportation, which is also an important aspect in collaborative vehicle routing. Because of this very recent survey, we can keep the cost allocation part short. 

Given the high volume of recent literature on collaborative vehicle routing it is now appropriate to provide a review on the state of knowledge. The contribution of our survey is threefold:

\begin{enumerate}
\item We also consider centralized collaborative planning.
\item We survey the literature of the last few years.
\item We give a new and broader classification of articles.
\end{enumerate}

The remainder of our survey is organized as follows. The research methodology used is described in Section~\ref{meth}. Classifications and definitions are provided in Section~\ref{class}. Centralized collaborative planning is surveyed in Section~\ref{central}. Sections~\ref{dec1} and \ref{dec2} give overviews on decentralized planning with and without auctions, respectively. Each section closes with a discussion on future research directions. A summarizing conclusion is given in Section~\ref{concl}.
\section{\textcolor[rgb]{0,0,0}{Research methodology}}\label{meth}
\textcolor[rgb]{0,0,0}{In our review, we focus on studies where \textit{operations research} models and solution techniques are applied. Pure empirical studies, not focusing on the operational planning problems, are not considered. However, readers interested in these empirical studies are referred to, e.g., \cite{Cruijssen2007}, \cite{Lydeka07}, \cite{ballot10}, \cite{schmoltzi2011}.
We also do not consider studies, where the main focus is on general design of coalitions rather, while the transportation planning problems of collaborators are neglected (e.g. \citealp{Tate96, Verstrepen09, Voruganti11, audy12, Guajardo2015ejor}). Regarding application fields, we limit our survey to studies on road transportation. Collaborations in rail (e.g. \citealp{Kuo08}), naval (e.g. \citealp{Agarwal2010}), and air (e.g. \citealp{Ankersmit2014}) transportation lead to interesting planning problems, but these studies do not fit within the scope of this survey. We initially did not omit any study based on its type (primary, secondary; journal articles, proceedings, etc.) or its year of publication.} 

\textcolor[rgb]{0,0,0}{With this scope in mind, we searched library databases, where the major journals in operations research, operations management, management science etc. are covered. Used search terms were basically combinations of}
\begin{itemize} 
\item{\textcolor[rgb]{0,0,0}{"collaboration", "cooperation", "coalition", "alliance" and}}
\item{\textcolor[rgb]{0,0,0}{"transportation", "routing", "logistics", "freight", "carrier", "shipper".}}
\end{itemize}

\textcolor[rgb]{0,0,0}{By this, we identified a first set of relevant studies. At this point we decided to only consider journal publications, in order to survey a reasonable number of articles. For each of the remaining papers we screened the reference list and added articles that had not been find in the first step. Finally, we applied a descendancy approach, which goes from old information to new information, by screening all articles that cite one of the papers that we found relevant. We proceeded until we converged to a final set of publications.}

\section{\textcolor[rgb]{0,0,0}{Classifications and definitions}}\label{class}
\begin{figure}[htbp]{
\begin{center}
%\fbox{
\includegraphics[width=0.9\textwidth,trim=3cm 17cm 2cm 2.5cm]{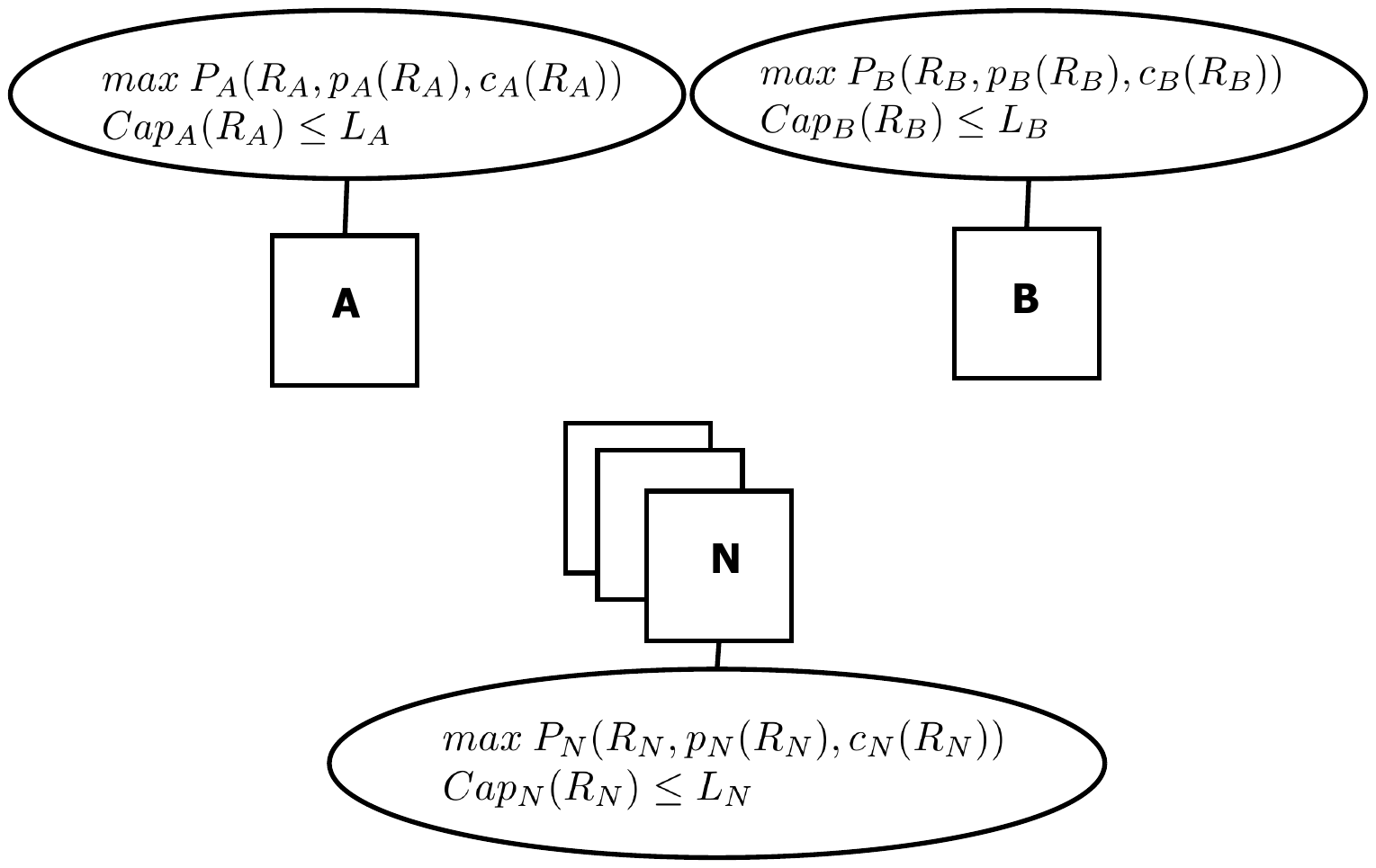}%}
\caption{\textcolor[rgb]{0,0,0}{Generalized illustration of a non-collaborative setting with profit ($P_i$) maximizing participants $i$, $i\in  (A,...,N)$, with limited capacity $L_i$. $Cap_i$ is the capacity usage.}}\label{fig:nocollab}
\end{center}}
\end{figure}

\begin{figure}[htbp]{
\begin{center}
%\fbox{
\includegraphics[width=0.6\textwidth,trim=3cm 16cm 2cm 2.5cm]{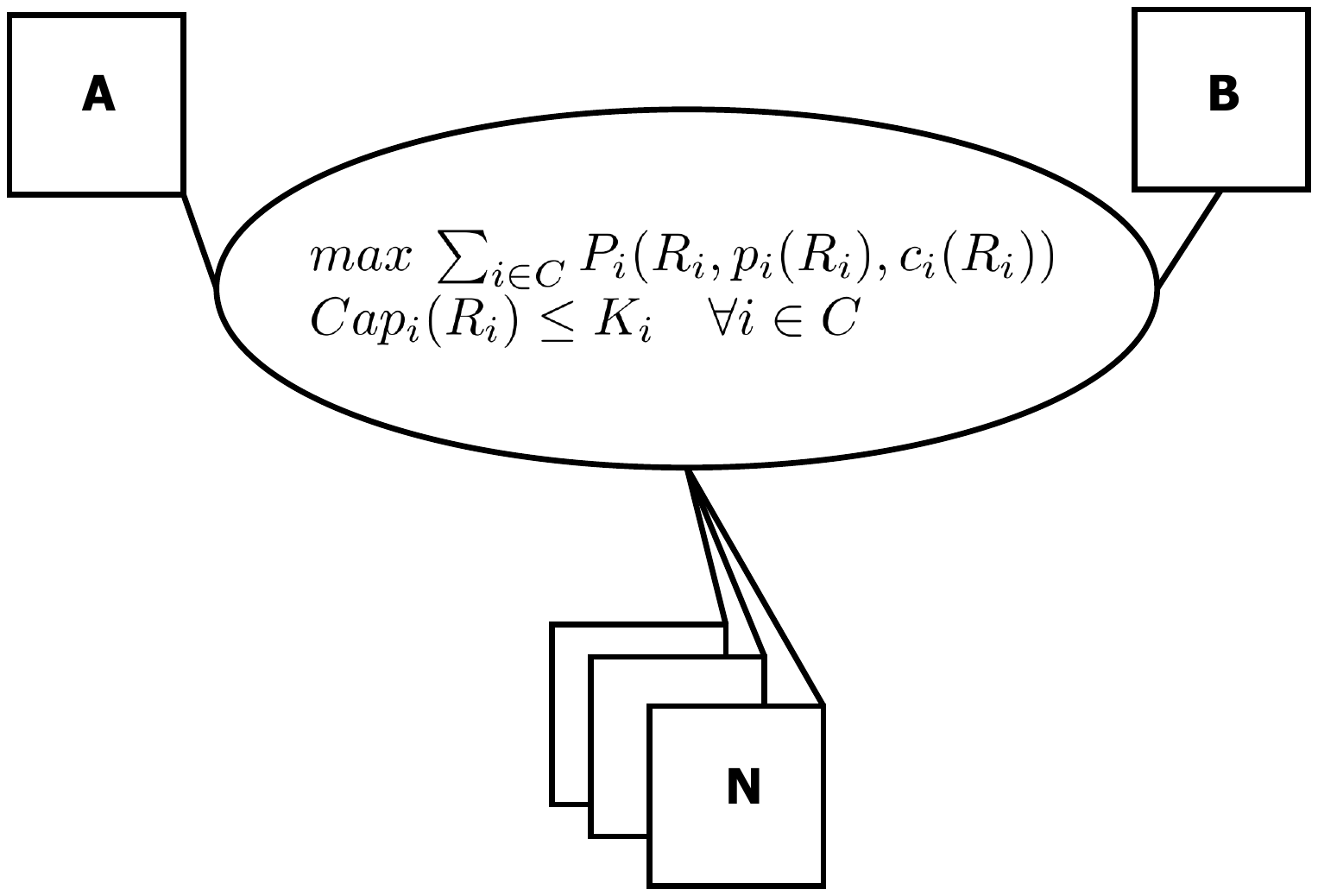}%}
\caption{\textcolor[rgb]{0,0,0}{Generalized illustration of centralized planning with profit ($P_i$) maximizing participants $i$, $i\in  (A,...,N)$, with limited capacity $L_i$. $Cap_i$ is the capacity usage.}}\label{fig:central}
\end{center}}
\end{figure}

\begin{figure}[htbp]{
\begin{center}
%\fbox{
\includegraphics[width=1.1\textwidth,trim=3cm 17cm 2cm 2.5cm]{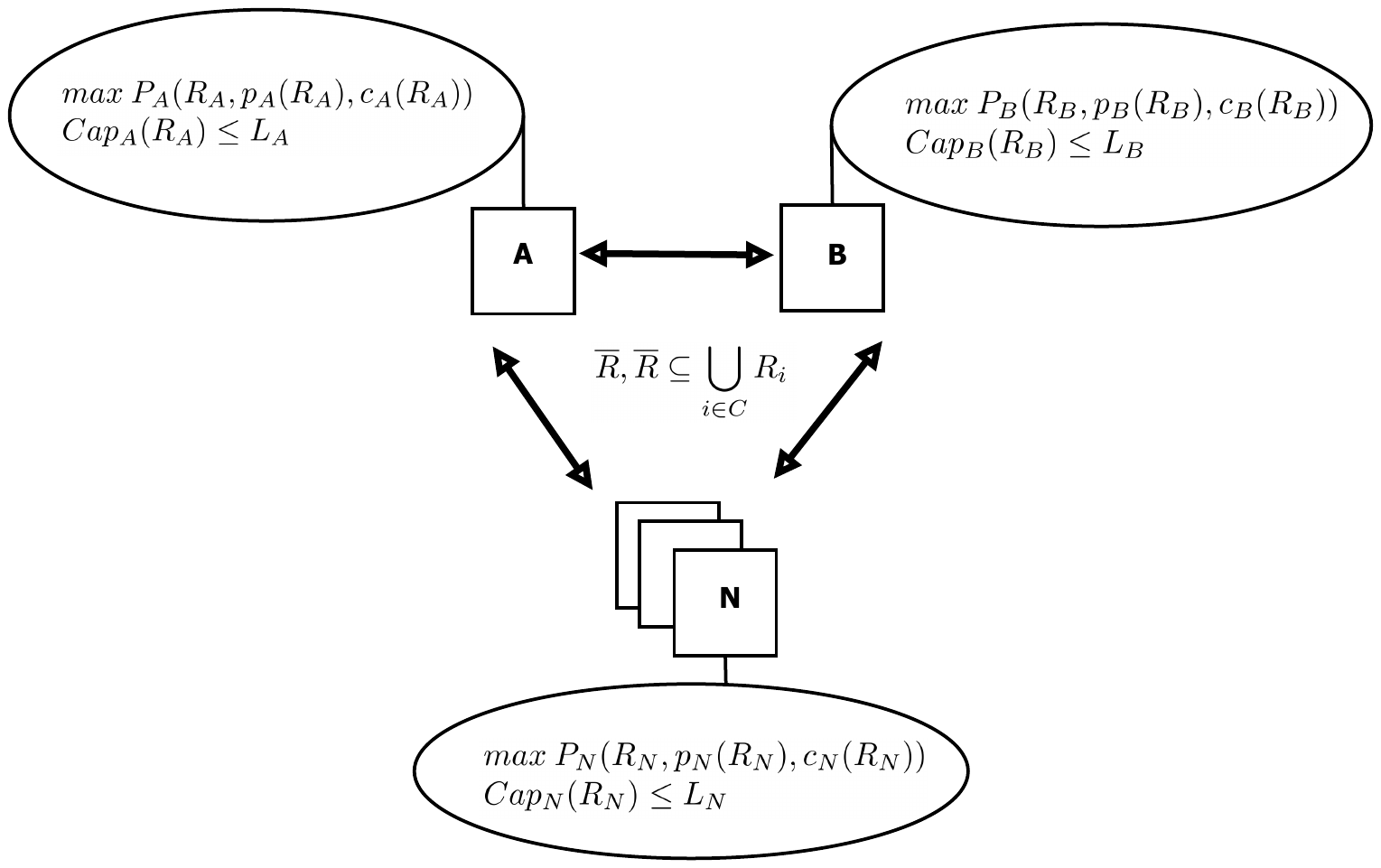}%}
\caption{\textcolor[rgb]{0,0,0}{Generalized illustration of decentralized planning with profit ($P_i$) maximizing participants $i$, $i\in  (A,...,N)$, with limited capacity $L_i$. $Cap_i$ is the capacity usage. $\bar{R}$ is the subset of requests that have been offered for exchange.}}\label{fig:decentral}
\end{center}}
\end{figure}

\textcolor[rgb]{0,0,0}{In our review, we distinguish between centralized and decentralized collaborative planning. In Figures~\ref{fig:nocollab}-~\ref{fig:decentral}, we provide a generalized illustrations of collaborative and non-collaborative settings. In the non-collaborative setting (Figure\ref{fig:nocollab}), each participant $i$, $i\in (A,...,N)$ maximizes his individual profit $P_i$. This profit depends on his set of requests $R_i$, the payments $p_i(R_i)$ that he gets for his requests $R_i$, and his costs $c_i(R_i)$. The capacity usage $Cap_i$ of a participant $i$ is limited by his available capacity $L_i$.}

\textcolor[rgb]{0,0,0}{In case of centralized planning (Figure~\ref{fig:central}), the total profit is maximized jointly (which is denoted as \textit{ideal model} by \citealp{Schneeweiss03}).} 

\textcolor[rgb]{0,0,0}{In decentralized settings (Figure~\ref{fig:decentral}), collaborators agree on a mechanism for exchanging subsets of their requests by revealing no or only limited information. $\bar{R}$ is the set of requests that have been offered for exchange. In our study, the research stream on decentralized planning is further split up in non-auction-based and auction-based studies.
}

All papers are classified based on the model formulation of the routing problem used. Here, the following categories can be identified. 
\textcolor[rgb]{0,0,0}{\begin{itemize}
\item \textit{Vehicle routing problems} (VRP), which give the optimal sets of routes for fleets of vehicles in order to visit a given set of customers (e.g. \citealp{Gendreau2008vrp}).
\item \textit{Arc routing problems} (ARP), which assume that customers are located on arcs that have to be traversed. The capacitated ARP is the arc-based counterpart to the node-based VRP (e.g. \citealp{Wohlk2008}).
\item \textit{Inventory routing problems} (IRP), which combine VRP with inventory management (e.g. \citealp{Bertazzi2008}).
\item \textit{Lane covering problems} (LCP), which aim at finding a set of tours covering all lanes with the objective of minimizing the total travel cost (e.g. \citealp{Ghiani08}). A lane is the connection between the pickup and the delivery node of a full truckload (FTL) request.
\item \textit{Minimum cost flow problems} (MCFP), which aim at sending goods through a network in the cheapest possible way (e.g. \citealp{klein67}).
\item \textit{Assignment problems} (AP), where vehicles are assigned to requests, such that total costs are minimized (e.g. \citealp{Munkres}).
\end{itemize}}

We also indicate whether models assume customers to have delivery \textit{time windows} (TW) or not. Another frequently appearing extension are \textit{pickup and delivery} (PD) requests. This means that goods have to be picked up at some node and to be delivered to another node, where pickup and delivery locations do not necessarily coincide with a depot. 

\begin{figure}[htbp]{
\begin{center}
%\fbox{
\includegraphics[width=0.8\textwidth,trim=3cm 4.5cm 2cm 2.5cm]{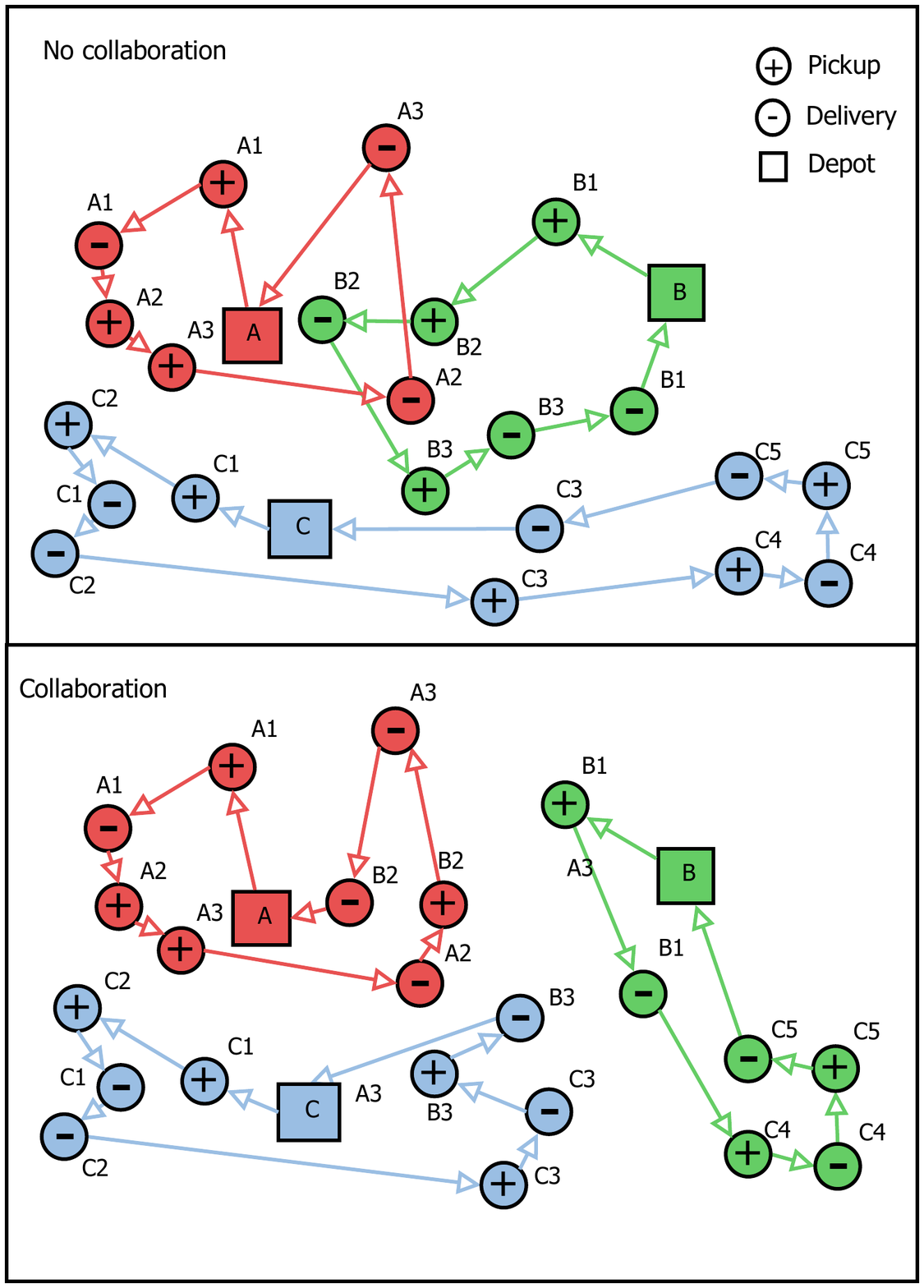}%}
\caption{Example for non-collaborative and collaborative vehicle routes of three LTL carriers with pickup and delivery requests (\citealp{GanstererIfors}).}\label{fig:ltl}
\end{center}}
\end{figure}

Articles can be further classified based on the investigated types of shipment. These can either be \textit{FTL} or \textit{less than truckload} (LTL). FTL can of course be seen as a special case of LTL, where the size of customer orders is equal to the vehicle's capacity. Hence, LTL models are applicable for FTL settings as well. However, FTL is often used in the transportation of a single product, whereas LTL is usually used to transport multiple products in small volumes from depots to customers or from customers to customers. A typical application area for LTL is parcel delivery (\citealp{DaiChen12, Parragh2008}). Figure~\ref{fig:ltl} shows an example of non-collaborative and of collaborative vehicle routes of LTL carriers.  

\begin{figure}[htbp]{
\begin{center}
%\fbox{
\includegraphics[width=0.6\textwidth,trim=3cm 3cm 2cm 2.5cm]{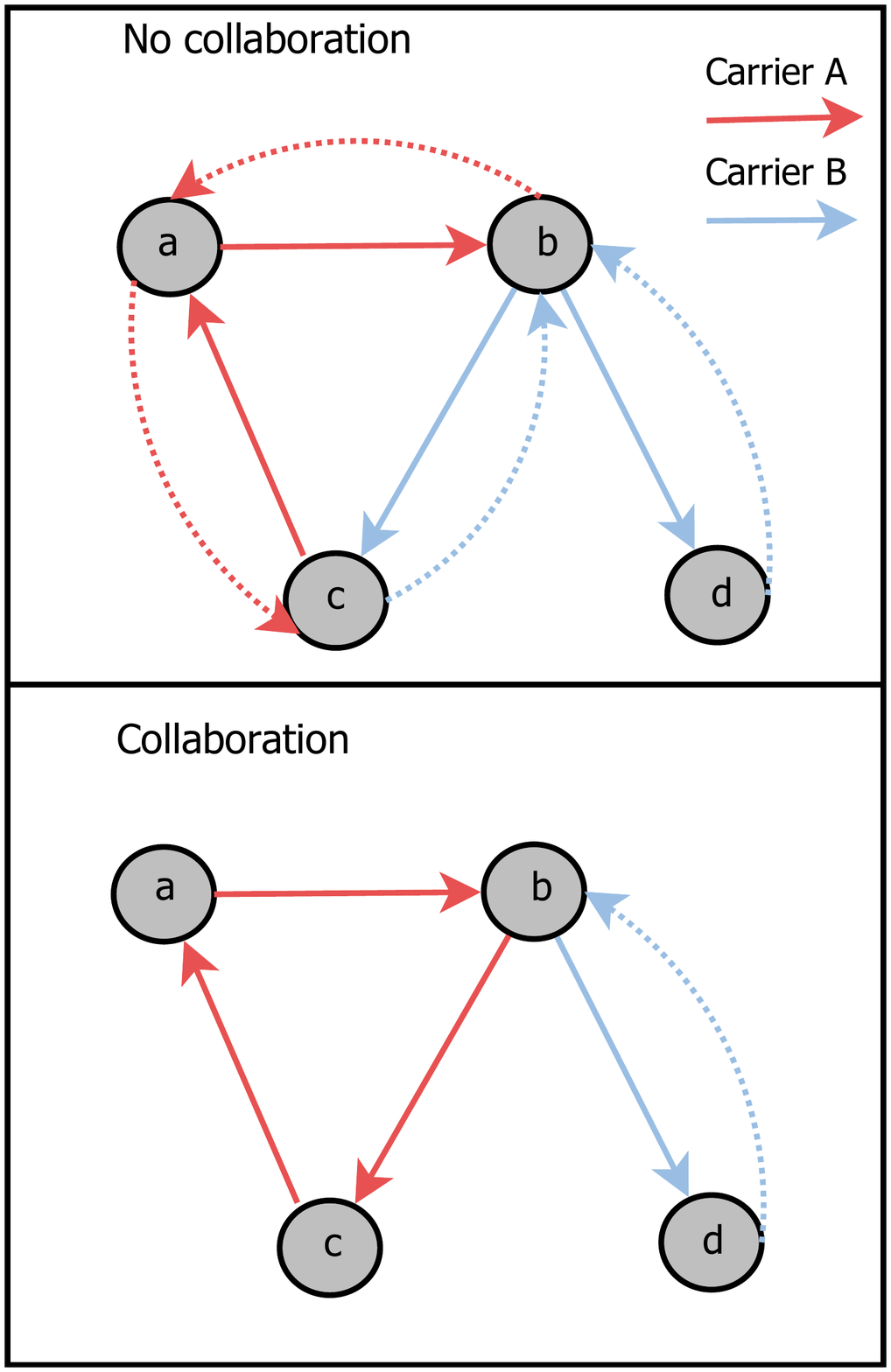}%}
\caption{Example for non-collaborative and collaborative vehicle routes of two FTL carriers. Dotted arcs are empty return trips (\citealp{Adenso2014})} \label{fig:ftl}
\end{center}}
\end{figure}

An example of collaboration between FTL carriers is displayed in Figure~\ref{fig:ftl}. Carrier A drives from customer c to customer a, and from customer a to customer b, while carrier B is serving lanes b-c and b-d. By collaboration the carriers can avoid three empty return trips (\citealp{Adenso2014}). It should be mentioned that b-c can of course also be used by carrier A in a non-collaborative setting. In this case this carrier has only one empty return trip. However, the collaboration with carrier B still improves the situation of carrier A.

\textcolor[rgb]{0,0,0}{Players in transportation collaborations might be \textit{carriers} (also denoted as freight forwarders, logistics service providers, or third party logistics providers) and \textit{shippers}. Carriers are assumed to be the owners and operators of transportation equipment, while shippers own or supply the shipments. Joint routing planning is typically assumed to be done by carriers. When shippers consider collaboration, they identify attractive bundles of lanes, helping carriers to reduce empty trips in return of better rates (\citealp{ErgunShipper07}).}

We observe that the huge majority of papers focuses on carrier-related cooperations. We agree with \cite{Cruijssen07joint}, that from the planning perspective it does not matter whether carriers or shippers are in charge of the process. However, in decentralized settings the issue of information asymmetries has to be taken into account. Shippers and carriers typically do not having the same level of information. We therefore find it useful to distinguish whether carriers or shippers are the players in a collaboration.  

In the papers surveyed, a variety of both, exact and heuristic solution methodologies are represented. An overview on these methodologies is given in Figure~\ref{fig:methods}. 

\begin{landscape}
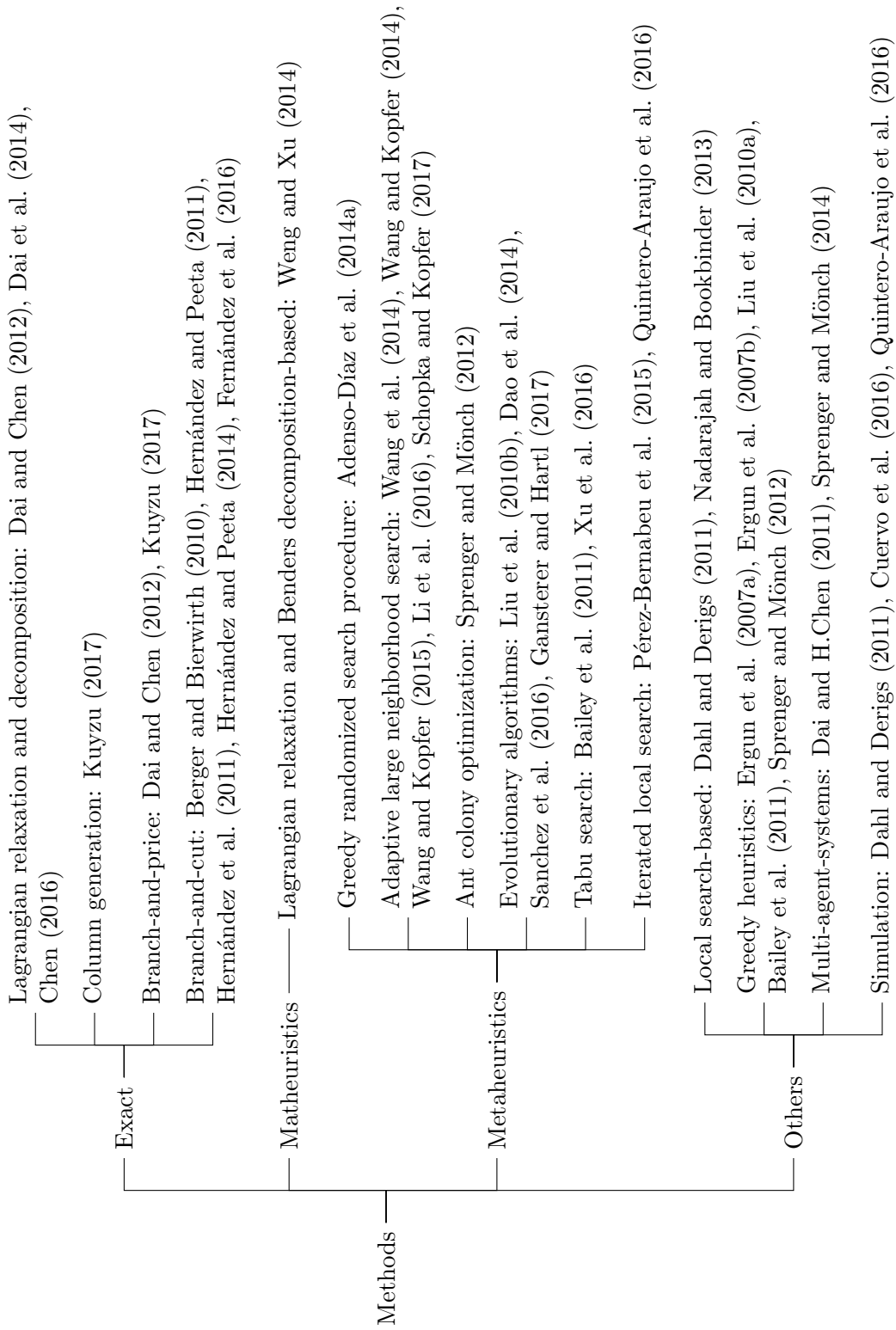
\begin{figure}

\centering
\newlength\treeheight
\setlength{\treeheight}{\textheight}
\small
\begin{tikzpicture}[grow=right, level distance=1cm,
  anchor=west,parent anchor=east , growth parent anchor=east,
  every node/.style={},
  level 1/.style={sibling distance=\treeheight/4},
  level 2/.style={sibling distance=\treeheight/15},
  %level 3/.style={sibling distance=\treeheight/15},
	%level 4/.style={sibling distance=\treeheight/16},
	]\node {Methods}[align=left,edge from parent fork right,grow=east]

child[sibling distance=42mm] {node {\textcolor[rgb]{0,0,0}{Others}}
child {node {\textcolor[rgb]{0,0,0}{Simulation: \cite{Dahl2011}, \cite{Cuervo16}, \cite{Quintero2016}}}}
child {node {\textcolor[rgb]{0,0,0}{Multi-agent-systems: \cite{Dai2011}, \cite{sprengerDSS}}}}
%child {node {Heuristics}
child {node {\textcolor[rgb]{0,0,0}{Greedy heuristics: \cite{Ergun07TS}, \cite{ErgunShipper07}, \cite{Liu2010},} \\ \textcolor[rgb]{0,0,0}{\cite{Bailey11}, \cite{sprenger2012}}}}
child {node {\textcolor[rgb]{0,0,0}{Local search-based: \cite{Dahl2011}, \cite{Nadarajah2013}}}}}
child  {node {Metaheuristics}
child {node {Iterated local search: \cite{Perez2015}, \cite{Quintero2016}}}
child {node {Tabu search: \cite{Bailey11}, \cite{Xu16}}}
child {node {Evolutionary algorithms: \cite{Liu2010task}, \cite{Dao2014}, \\ \cite{Sanchez2016}, \cite{Gansterer2017}}}
child {node {Ant colony optimization: \cite{sprenger2012}}}
child {node {Adaptive large neighborhood search: \cite{WangKG2014}, \cite{Wang2014}, \\ \cite{WangKopf2015}, \cite{Li2016}, \cite{Schopka2017}}}
child {node {Greedy randomized search procedure: \cite{Adenso2014}}}}
child [sibling distance=30mm]{node {Matheuristics}
child {node {Lagrangian relaxation and Benders decomposition-based: \cite{Weng2014}}}}
child [sibling distance=27mm]{node {Exact}
child {node {Branch-and-cut: \cite{Berger2010}, \cite{HernandezCentr11}, \\ \cite{Hernandez11TRE}, \cite{Hernandez14Trat}, \cite{Fernandez16}}}
child {node {Branch-and-price: \cite{DaiChen12}, \cite{Kuyzu17}}}
child {node {Column generation: \cite{Kuyzu17}}}
child {node {Lagrangian relaxation and decomposition: \cite{DaiChen12}, \cite{Dai2014}, \\ \cite{Chen2016}}}}

;
\end{tikzpicture}
\caption{Overview on solution methodologies.}
\label{fig:methods}
\end{figure}
\end{landscape}

\textcolor[rgb]{0,0,0}{An important aspect of collaborative operations is how to share the gained benefits. It is shown in \cite{Guajardosurvey} that most problems in collaborative transportation use sharing methods based on \textit{cooperative game theory}. The authors identify more than 40 different methods, which they categorize as \textit{traditional} or \textit{ad hoc} concepts. However, they show that in the huge majority of studies, one of the following three methods is used:}
\begin{itemize}
\item \textcolor[rgb]{0,0,0}{the well-known \textit{Shapley value} (\citealp{shapley}), which is generally the most applied method (e.g. \citealp{Kimms2016, Vanovermeire2014cost, engevall}).}
\item \textcolor[rgb]{0,0,0}{\textit{proportional methods}, where each carrier $j$ gets a share $\alpha_j$ of the total profit (e.g. \citealp{oezener13, Berger2010, FGJR2013}}
\item \textcolor[rgb]{0,0,0}{the \textit{nucleolus method} initially defined by \cite{schmeidler} (e.g. \citealp{Agarwal2010, Guajardo2015nuc, Göthe-Lundgren1996}).}
\end{itemize}

\textcolor[rgb]{0,0,0}{In the literature, there are basically two streams of research: (i) articles that focus on the transportation problems, while profit sharing is not taken into account, and (ii) articles dealing with profit sharing, while the transportation problem is neglected. There are only a few studies (for instance \citealp{KrajLap08}), where both aspects are combined. Thus, our survey aims at building bridges between these two worlds. In our sections on future research, we provide several suggestions how profit sharing aspects should be integrated into the collaborative planning processes.}

\section{Centralized collaborative planning}
\label{central}
 
If collaborative decisions are made by a central authority having full information, this is referred to as \textit{centralized collaborative planning}. An example for such a central authority might be an online platform providing services for collaborative decision making (\citealp{DaiChen12}). It is obvious that under full information, the decision maker has to tackle a standard optimization problem, since the collaborative aspect is diminished by information disclosure. Thus, each transportation planning problem might be interpreted as a collaborative transportation planning problem that a decision maker with full information has to solve. However, to have in this review a reasonable scope, we only survey studies that contribute to the collaborative aspect of transportation planning. An example for such a contribution is, for instance, given in \cite{WangKG2014}. In this study the central authority does not have full power to find an optimal solution by simply exchanging requests. It rather has to decide on the degree of collaboration taking both outsourcing and request exchange into consideration. 

It can be observed that that there are two streams of research in this area:
\begin{itemize}
\item the assessment of the potential benefits of centralized collaborative planning versus non-cooperative settings. In non-cooperative settings players do not show any kind of collaborative efforts. We refer to this research stream as \textit{Collaboration Gain Assessment (CGA)}.
\item innovative models or innovative solutions approaches for centralized collaborative planning. These studies will be denoted as \textit{Methodological Contributions (MC)}.   
\end{itemize}

Table~\ref{tab:central} gives an overview on studies contributing to centralized collaborative planning. We classify them according to the categories given above, and highlight their main characteristics.   

\begin{landscape}
\begin{table} 
\caption{References and basic characteristics for collaboration with full information.}
\smallskip
\label{tab:central}  
\begin{tabular}{lccccccc}
\hline\noalign{\smallskip} 
Reference&Focus&Shipper&Carrier&Model&Shipment&TW&PD\\
\hline
\cite{Adenso2014}&CGA&x&&VRP&FTL&&x\\ %; GRASP;shipper
\cite{Buijs16}&MC&&x&VRP&LTL&x&x\\%carrier (LSP); Decomposition heuristic
\cite{Cruijssen07joint}&CGA&x&x&VRP&LTL&x&\\%sagt, dass es egal ist wer das macht (carrier (LSP) oder shipper, aber wir könnten vielleicht sagen, dass joint route planning eher was für carrier ist. sagt er das sogar?); heuristic
\cite{DaiChen12}&MC&&x&VRP&LTL&&x\\%carrier; Lagrangian relaxation and decomposition
\cite{ErgunShipper07}&MC&x&&LCP&FTL&&x\\%Shipper; Greedy algorithm
\cite{Fernandez16}&MC&&x&ARP&LTL&&\\%Carrier; B-and-C
\cite{HernandezCentr11}&MC&&x&MCFP&LTL&&\\ %Carrier; B-and-C
\cite{KrajLap08}&CGA&&x&VRP&LTL&x&x\\ %Carrier; B-and-C
\cite{Kuyzu17}&MC&x&&LCP&FTL&&x\\%Shipper, column generation and branch-and-price
\cite{Lin08}&CGA&&x&VRP&LTL&x&x\\%carriers, decomposition heuristic
\cite{Liu2010}&MC&&x&ARP&FTL&&x\\%Carrier; greedy algorithm
\cite{Montoya16}&CGA&&x&VRP&LTL&&\\%Carrier, heurisitic
\cite{Nadarajah2013}&MC&&x&VRP&LTL&x&\\%carrier, local searh (math) heuristic
\cite{Perez2015}&CGA&&x&VRP&LTL&&\\%carrier, iLS
\cite{Quintero2016}&CGA&&x&VRP&LTL&&\\%carrier, Simheuristic with ILS and biased randomization
\cite{Sanchez2016}&CGA&&x&VRP&LTL&x&\\%carrier; MH: scatter search
\cite{Soysal2016}&CGA&&x&IRP&LTL&&\\%carrier (3pl)Exact (Cplex)
\cite{sprenger2012}&CGA&&x&VRP&LTL&x&\\%carrier, decomposition heuristic, greedy heuristic und acs
\cite{WangKG2014}&MC&&x&VRP&LTL&x&x\\%carrier (freigth forwarder), heuristic and ALNS
\cite{Weng2014}&MC&&x&ARP&LTL&&\\%carriers, heuristics based on LR and BD
\cite{Yilmaz2012}&CGA&x&&AP&LTL&x&x\\%shippers, simulation of a markov decision model
\noalign{\smallskip}\hline
\end{tabular}
\end{table} 
\end{landscape}
  
\subsection{Collaboration gain assessment}

One of the first studies to systematically assess the potentials of collaborative vehicle routing was presented by \cite{Cruijssen07joint}. The authors consider a system with multiple companies, each having a separate set of distribution orders. Goods are picked up at a single distribution center and delivered to customer sites. Both, a non-cooperative setting, where each company solves the planning problem independently, and a cooperative setting, where routes are planned jointly are investigated. It is shown that joint route planning can achieve synergy values of up to 30\%. 

Many other studies confirm the observation of \cite{Cruijssen07joint}, that centralized collaborative planning has the potential to improve total profits by around 20-30\% of the non-cooperative solution (e.g. \citealp{Montoya16, Soysal2016}). A real-world setting, where a local courier service of a multi-national logistics company is investigated by \cite{Lin08}. It is shown that the cooperative strategy, where courier routes are planned jointly, outperforms the non-cooperative setting by up to 20\% of travel cost.    

Joint route planning is generally considered to be done by carriers (e.g. \citealp{DaiChen12, Buijs16, Liu2010}), but also shippers can be involved in joint route planning, as long as they have direct control over the flows of goods (\citealp{Cruijssen07joint}). However, merging FTL lanes is mostly assumed to be done by shippers (e.g. \citealp{Adenso2014,ErgunShipper07, Kuyzu17}). Here again one might argue, that also carriers can be involved in this type of horizontal collaboration, as it is considered by, e.g., \cite{Liu2010}.  

Horizontal collaborations not only follow economical but also ecological goals like reduced road congestion, noise pollution, and emissions of harmful substances. Thus, public authorities are encouraging companies to collaborate. The city of Zurich, for instance, is funding a research project aiming at improved cooperation between different transport companies by an IT-based collaboration platform (\citealp{zhaw}). In this spirit, \cite{Montoya16} quantify the effect of collaborative routing in the field of city logistics. In order to solve real-world instances from the city of Bogot\'{a}, the centralized problem is decomposed into an assignment and a routing part. By this, the non-cooperative solution can be improved by 25.6\% of the travel distance. 

Many other recent studies account for ecological aspects. \cite{Perez2015}, for instance, examine different VRP scenarios and show that cooperations can contribute to a noticeable reduction of expected travel costs as well as of greenhouse gas emissions. The VRP with time windows (VRPTW) and with carbon footprint as a constraint is proposed by \cite{Sanchez2016}. Using this model, the reduction of carbon emissions in a collaborative setting, where different companies pool resources, is investigated. The authors find that the total greenhouse gas emissions can be reduced by 60\%, while cost savings were nearly 55\%. \cite{Soysal2016} model and analyse the IRP in a collaborative environment, which accounts for perishability, energy use ($CO_2$ emissions), and demand uncertainty. According to their experiments,
the cost benefit from cooperation varies in a range of about 4–-24\%, while the aggregated total emission benefit varies in a range of about 8–-33\%. 

Collaboration potentials in stochastic systems has also been assessed by \cite{sprenger2012}. The authors investigate a real-world scenario found in the German food industry, where products are sent from manufacturers to customers via intermediate distribution centers. They are the first to show that the cooperative strategy clearly outperforms the non-cooperative algorithms in a dynamic and stochastic logistics system. A large-scale VRPTW is obtained for the delivery of the orders, capacity constraints, maximum operating times for the vehicles, and outsourcing options. This problem is decomposed into rich VRP sub problems and solved by an algorithm based on ant colony systems. The proposed heuristics is tested in a rolling horizon setting using discrete event simulation.

\cite{Quintero2016} discuss the potential benefits of collaborations in supply chains with stochastic demands. A simheuristic approach is used to compare cooperative and non-cooperative scenarios. The authors find costs reduction around 4\% with values rising up to 7.3\%. 

\cite{Yilmaz2012} study the collaboration of small shippers in the presence of uncertainty. This problem focuses on markets where shippers have random transportation requests for small-volume shipments. The AP, where the coalition decides where to assign an arriving shipper and when to dispatch a vehicle, is proposed and modeled as a Markov decision process. Its performance is compared to a naive and a myopic strategy. The authors find, for instance, that when it is costly to pickup and deliver the shipments to
consolidation points, the naive policy is outperformed by the policy of the coalition. 

While the majority of papers finds that horizontal collaborations can improve the non-cooperative solution by around 20-30\%, some authors report collaboration profits outside of this range (e.g. \citealp{KrajLap08, Sanchez2016, Quintero2016}). \cite{Adenso2014} contribute to this issue by investigating the impact of coalition sizes. Their computational experiments show that benefits are marginally decreasing with the size of the partnership. 

\subsection{Methodological contributions}

In centralized planning problems, there are several decisions that have to be taken. Typically, not only the routing but also the assignment of customers to depots has to be considered. In order to approximate optimal solutions even for large real world instances, many authors propose decomposition strategies (e.g. \citealp{DaiChen12, Nadarajah2013, Buijs16}). 

%\cite{DaiChen12} propose a decomposition approach for a carriers’ collaborative LTL transportation planning problem with pickup and deliveries. Their method consists of two steps. Firstly, a mixed integer programming model, which is a generalisation of the LCP, is proposed. Secondly, a set of feasible vehicle tours is constructed.  

%\cite{Nadarajah2013} also present a two stage framework for LTL carrier collaboration. The first stage refers to collaboration between multiple carriers at the entrance to a city, which can be formulated as a VRPTW. The second stage involves collaboration between carriers at transshipment facilities. 

%\cite{Buijs16} study the collaboration between two business units of Fritom, a Dutch logistics service provider, and propose alternatives to improve its collaborative transport planning. They introduce the generalized pickup and delivery problem, which relaxes the common constraints that a load must be transported from its origin to its destination using one vehicle within a single planning period. The authors show different decomposition approaches, which are necessary to solve the real-world instances. 

While a popular assumption is that in horizontal collaborations the VRP is the underlying planning problem, also collaborative ARP or MCFP have been investigated. \cite{Fernandez16} introduce the \textit{collaboration uncapacitated ARP}. This yields to a \textit{profitable ARP}, where carriers have customers that they are not willing or allowed to share, and others that can be exchanged with collaborators. The model is formulated as integer linear program and solved through a branch-and-cut algorithm. The optimal \textit{hub routing problem} of merged tasks is investigated by \cite{Weng2014}. This problem allows all requests to pass up to two hubs within limited distance. The underlying problem is formulated as multi-depot ARP. Solutions are generated using two heuristics based on Lagrangian relaxation and Benders decomposition. The \textit{time-dependent centralized multiple carrier collaboration problem} is introduced by \cite{HernandezCentr11}. The authors assume a setting where carriers either provide or consume collaborative capacity. Capacities are time-dependent but known a priori, and demand is fixed. The problem is modeled as a binary multi-commodity MCFP and solved using a branch-and-cut algorithm. \cite{Liu2010} define the \textit{multi-depot capacitated ARP} aiming for a solution with minimized empty movements of truckload carriers. A two-phase greedy algorithm is presented to solve practical large-scale problems. 

Not only carriers, but also shippers can conduct horizontal collaborations. In this case, they jointly identify sets of lanes that can be submitted to a carrier as attractive bundles. The goal is to offer tours with little or no asset repositioning to carriers. In return, they can get more favorable rates from the carriers. \cite{ErgunShipper07} define the \textit{shipper collaboration problem}, which is formulated as LCP. Solutions are generated by a greedy algorithm. In \cite{Kuyzu17} the LCP model is extended by a constraint on the number of partners with whom the collaborative tours must be coordinated. Column generation and Branch-and-price approaches are developed for the solution of the resulting LCP variant. 

In collaborative vehicle routing, typically horizontal cooperations are considered. These refer to collaborative practices among companies acting at the same levels in a market (\citealp{CruijsenSurv}). Vertical cooperation on the other hand, indicate hierarchical relationships, meaning that one player is the client of the other. \cite{WangKG2014} were the first to present a combination of horizontal and vertical cooperation. They extend the pickup and delivery problem with time windows (PDPTW) to a combination of integrated (vertical) and collaborative (horizontal) transportation planning, where both subcontracting and collaborative request exchange are taken into account. \textcolor[rgb]{0,0,0}{The centralized planning problem is solved using adaptive large neighborhood search (ALNS) and an ALNS-based iterative heuristic}.

\subsection{Future research directions}
\textcolor[rgb]{0,0,0}{The area of CGA has been extensively researched. The cost advantages of centralized collaborations have been quantified in several studies, most of them finding potential benefits of 20-30\%. Also ecological goals, like reduction of emissions, have been taken into account. However, most of these studies assume deterministic scenarios. Literature assessing collaboration potentials, when the central authority faces uncertainties, is scarce. Also collaboration gains in more complex, e.g. multi-modal, multi-depot transportation systems have yet to be investigated.}  

\textcolor[rgb]{0,0,0}{Centralized authorities typically face huge and highly complex optimization problems, since they have to plan operations for several interconnected fleets. Thus, sophisticated solution techniques are required. There is a vast field of problems and methods that have not been investigated so far from a collaborative perspective. It could, for instance, be investigated, how a central authority exchanges requests among collaborators, while trying not to redistribute too much. This would lead to a 2-objective problem, which minimizes (i) total cost and (ii) deviation from the decentralized solution. A related question is how the central authority can motivate participants to reveal their data. These incentives might be provided by using smart profit sharing mechanisms or, e.g., side payments. To answer these questions, studies investigating how much information has to be revealed in order to achieve reasonable collaboration profits would be helpful.}
\textcolor[rgb]{0,0,0}{
Finally, since central decision makers face huge optimization problems, the application of solution methods for large scale VRP (e.g. \citealp{Braysy2007}) are supposed to further improve solution quality. For this purpose, advanced processing methods like parallel computing should be taken into account (\citealp{GHIANI20031}). Also machine learning concepts might be valuable tools to, e.g., tune parameters (\citealp{Biratarri}) or to automatically identify problem structures in collaborative settings.  
}

\section{Decentralized planning without auctions}
\label{dec1}

If players are not willing to give full information to a central planner, decentralized approaches are needed. In such a decentralized setting collaborators might cooperate individually or supported by a central authority, which does not have full information. 
Articles in this area contribute either to the issue of
\begin{itemize}
\item selecting appropriate collaboration partners, we refer to this as \textit{Partner Selection} (PS),
\item requests that should be offered to collaboration partners, which is referred to as \textit{Request Selection} (RS),
\item methods for exchanging requests, which is denoted \textit{Request Exchange} (RE).
\end{itemize}

In Table \ref{tab:decentral}, we categorize all papers dealing with decentralized non-auction-based approaches and give their main characteristics. 

\begin{landscape}
\begin{table} 
\caption{References and basic characteristics for non-auction-based decentralized collaboration}
\smallskip
\label{tab:decentral}  
\begin{tabular}{lccccccc}
\hline\noalign{\smallskip} 
Reference&Main contribution&Shipper&Carrier&Model&Shipment&TW&PD\\
\hline
\cite{Adenso14syn}&PS&&x&VRP&FTL&&x\\%carriers, heuristic
\cite{Bailey11}&RS&&x&VRP&FTL&&x\\%carriers, greedy heuristic and TS
\cite{Cuervo16}&PS&&x&VRP&LTL&&\\%Carriers, simulation
\cite{Dahl2011}&RE&&x&VRP&LTL&x&x\\%Carriers, simulation and lS heuristic
\cite{Dao2014}&PS&&x&AP&FTL&x&x\\%Carrier, GA
\cite{Ergun07TS}&RS&&x&LCP&FTL&x&x\\%carrier, greedy heuristic
\cite{Hernandez11TRE}&RS&&x&MCFP&LTL&&x\\%carrier, branch-and-cut
\cite{Hernandez14Trat}&RS&&x&MCFP&LTL&&x\\%carrier, branch-and-cut
\cite{Liu2010task}&RS&&x&ARP&FTL&&x\\%carrier, memetic algorithm
\cite{Ozener11}&RE&&x&LCP&FTL&&x\\%carrier, heuristics
\cite{sprengerDSS}&RE&&x&VRP&LTL&x&\\%Carrier, multi-agent-system
\cite{Wang2014}&RE&&x&VRP&LTL&x&x\\%carrier (freigth forwarder), heuristic and ALNS
\cite{WangKG2014}&RE&&x&VRP&LTL&x&x\\%carrier (freigth forwarder), heuristic and ALNS
\cite{WangKopf2015}&RE&&x&VRP&FTL&x&x\\%carrier (freigth forwarder), heuristic and ALNS
\hline
\noalign{\smallskip}\hline
\end{tabular}
\end{table} 
\end{landscape}

\subsection{Partner selection}
The benefit of collaborations of course depends on the partners that form the coalition and the characteristics of their operations.
Potential partners might have different requirements, which have to be considered in the joint operational plan (\citealp{Cuervo16}). Thus, \cite{Adenso14syn} propose an a priori index that could be used to roughly predict synergies between potential partners, without the necessity of solving any optimization model. This index is based on the transportation demands of the participating companies. However, average order size and the number of orders seem to be the most influential characteristic on the coalition’s profit  (\citealp{Cuervo16}). 

A model integrating partner selection and collaborative transportation scheduling is developed by \cite{Dao2014}. The mathematical model basically accounts for transportation times, costs, and capabilities of potential partners.  

\subsection{Request selection}
Carriers have to decide which of their requests should be offered to collaboration partners. Typically, carriers do not want to offer all their requests, but to keep some of them to be served with their private fleet. An intuitive solution would be to let the carriers solve a team orienteering problem, and put those requests into the pool, that do not appear in the optimal tour (Archetti et al., 2014). A combination of the request selection and the routing problem of collaborative truckload carriers is introduced by \cite{Liu2010task}. It is assumed that carriers receive different kinds of requests, which they have to allocate to either their internal fleet or to an external collaborative carrier. The objective is to make a selection of tasks and to route the private vehicles by minimizing a total cost function. The authors develop a memetic algorithm to solve the problem. However, \cite{Gansterer2016} show that in auction-based exchanges, the best request evaluation criteria take geographical aspects into account. It can be assumed, that these criteria are effective in non-auction-based settings as well. 

Of course, carriers not only decide on requests they want to offer, but on requests they want to acquire. Again, solving a team orienteering problem, which gives the set of valuable requests, would be an intuitive approach. However, this comes with a high computational effort since various different restrictions have to be considered. A request might be valuable for different players in the decentralized system. Thus, a coordinating authority has to find a feasible assignment of requests to carriers. An efficient method to reduce empty backhauls by adding pickup and delivery tasks of partners is proposed by \cite{Bailey11}. Two optimization models are developed, where one is formulated as an integer program, and the other is formulated as a mixed integer program. A greedy heuristic and tabu search, are used to solve these problems. A numerical analysis based on real-world freight data indicates that the percentage of cost savings can be as high as 27\%.

In \cite{Hernandez14Trat} a carrier seeks to collaborate with other carriers by acquiring capacity to service excess demand. Carriers first allocate requests to their private resources and then find the cost minimizing transport option for excess demand. The problem is addressed from a static perspective, formulated as a MCFP, and solved using a branch-and-cut algorithm. In \cite{Hernandez11TRE} dynamic capacities are assumed. 

In vertical collaborations, beneficial requests or tours have to be selected for collaboration partners. \cite{Ergun07TS} investigate vertical collaboration between shippers and carriers. The authors present optimization technology that can be used by shippers to identify beneficial tours with little truck repositioning. Timing considerations are a key focus of their study. The effectiveness of their algorithms is shown based on real-world data.

\subsection{Request exchange}
Once collaboration partners and requests have been selected, players have to decide on the exchange mechanism. Due to the inherent complexity, it is not common to exchange sets of unconnected requests, but parts of existing tours. This complexity can be overcome by auction-based systems. We refer to Section~\ref{dec2}, where auction-based mechanisms are discussed. However, in non-auctioned-based frameworks, it is reasonable to trade requests being packed in vehicle routes. \cite{Wang2014} propose such an route-based exchange mechanism. Carriers can iteratively generate and submit new routes based on the feedback information from an agent. This information is deduced from the dual values of a linear relaxation of a set partitioning problem. An extension including subcontracting is presented in \cite{WangKG2014}.
% The model is embedded into a framework, where the planning problem has to be solved by each member in a coalition. 

The problem becomes even more complex, if dynamics of carrier coalitions are considered. Two rolling horizon planning approaches, which yield considerably superior results than isolated planning, are proposed by \cite{WangKopf2015}. 

In the FTL market, typically lanes rather than requests are exchanged. \cite{Ozener11} consider settings in which several carriers collaborate by means of bilateral lane exchanges with and without side payments. 

With regard to applicability of request exchange mechanisms for real-world collaborations, some decision support systems have been developed. 
\cite{Dahl2011} present an empirical analysis of the effectiveness of a collaborative decision support system in an express
carrier network. They assume carriers to solve a dynamic PDPTW by a local search-based heuristic. They show that with the support of a
real-time decision support system based on an adequate compensation scheme, the network is able to perform close to the level obtainable by centralized planning. A decision support system for cooperative transportation planning where several manufacturing companies share their fleets to reduce transportation costs is presented by \cite{sprengerDSS}. 
%teilweise vertical, request exchange

\subsection{Future research directions}
Decentralized non-auction-based systems have advantages and disadvantages. They are generally assumed to be less complex than auction-based approaches, since there is, for instance, no need for a bidding procedure. This might be seen as an advantage, but this comes at the price that none of the players has structured information on the collaborators' preferences. This of course leads to relatively low collaboration profits. To overcome this drawback, there are attempts to still get some information by, for instance, doing multiple rounds of exchanges in order to approximate mutual preferences (e.g. \citealp{Wang2014}). An interesting research direction might be to compare the performance of such mechanisms with auction-based systems. Also the value of sharing information with collaboration partners has not been investigated so far. 

\section{Auction-based decentralized planning}
\label{dec2}

The decentralized exchange of requests can be organized through auctions (e.g. \citealp{Ledyard2002}), where collaborators submit requests to a common pool. Due to the necessity of a trading mechanism, auctions are generally supposed to be more complex than their conventional (i.e. non-auction-based) counterparts. However, auctions have more potential, since the trading mechanism can be used to indirectly share information of collaborators' preferences.   

In horizontal collaborations, auctions are used to exchange requests. Thus, collaborators typically have both the roles of buyers and of sellers. A central authority, which is in charge of coordinating the auction process, is called the auctioneer.

In combinatorial auctions, requests are not traded individually but are combined to bundles (\citealp{Pekec2003}). This is of particular importance in vehicle routing, where a request might not be attractive unless it is combined with other ones. An example for this can be seen in the upper part of Figure~\ref{fig:ltl}. Carrier B would probably not be interested in the individual requests C4 or C5, while the bundle (C4, C5) seems to be attractive. 

A bidding carrier receives the full bundle if the bidding price is accepted. If the bid is rejected, none of the items contained in the package is transferred to this carrier. This eliminates the risk of obtaining only a subset of requests which does not fit into the current request portfolio. Table \ref{tab:decauc} lists papers on auction-based decentralized planning, and highlights their basic characteristics.

%\begin{landscape}
\begin{table} 
\caption{References and basic characteristics for auction-based decentralized planning}
\smallskip
\small
\label{tab:decauc}  
\begin{tabular}{lcccccc}
\hline\noalign{\smallskip} 
Reference&Model&Shipper&Carrier&Shipment&TW&PD\\
\hline
\cite{Ackermann2011}&VRP&&x&LTL&&x\\%Carrier, heuristics
\cite{Berger2010}&VRP&&x&LTL&&x\\%Carrier, Branch-and-cut and exact (Cplex)
\cite{Chen2016}&VRP&&x&LTL&x&x\\%Carrier, Lagrangian Relaxation und decomposition
\cite{Dai2011}&VRP&&x&LTL&&x\\%Carrier und MAS
\cite{Dai2014}&VRP&&x&LTL&x&x\\%Carrier, Lagrangian Relaxation und decomposition
\cite{Gansterer2016}&VRP&&x&LTL&&x\\%Carrier, heuristics
\cite{Gansterer2017}&VRP&&x&LTL&&x\\%Carrier, heuristics, GA
\cite{Krajewska2006}&VRP&&x&LTL&x&x\\%Carrier, heuristics
\cite{Li2016}&VRP&&x&LTL&x&x\\%Carrier, ALNS
\cite{LiRong2015}&VRP&&x&FTL&&x\\%Carrier
\cite{Schopka2017}&VRP&&x&LTL&x&\\%Carrier, ALNS
\cite{Xu16}&LCP&&x&FTL&&x\\%Carrier, TS
\hline
\hline
\noalign{\smallskip}\hline
\end{tabular}
\end{table} 
%\end{landscape}

An early study on auction-based horizontal transportation collaboration is presented by \cite{Krajewska2006}. They are the first to design an auction-based exchange mechanism for collaborating carriers. \cite{Ackermann2011} discuss various goals for a combinatorial request exchange in freight logistics. They give a complete modeling proposal and show the crucial points when designing such a complex system. \cite{Berger2010} divide the auction process into 5 phases:

\begin{enumerate}
\item Carriers decide which requests to put into the auction pool.
\item The auctioneer generates bundles of requests and offers them to the carriers.
\item Carriers place their bids for the offered bundles.
\item \textit{Winner Determination Problem}: Auctioneer allocates bundles to carriers based on their bids.
\item Profit sharing: collected profits are distributed among the carriers.
\end{enumerate}

\begin{figure}[htbp]{
\begin{center}
%\fbox{
\includegraphics[width=0.8\textwidth,trim=3cm 17cm 3cm 2.5cm]{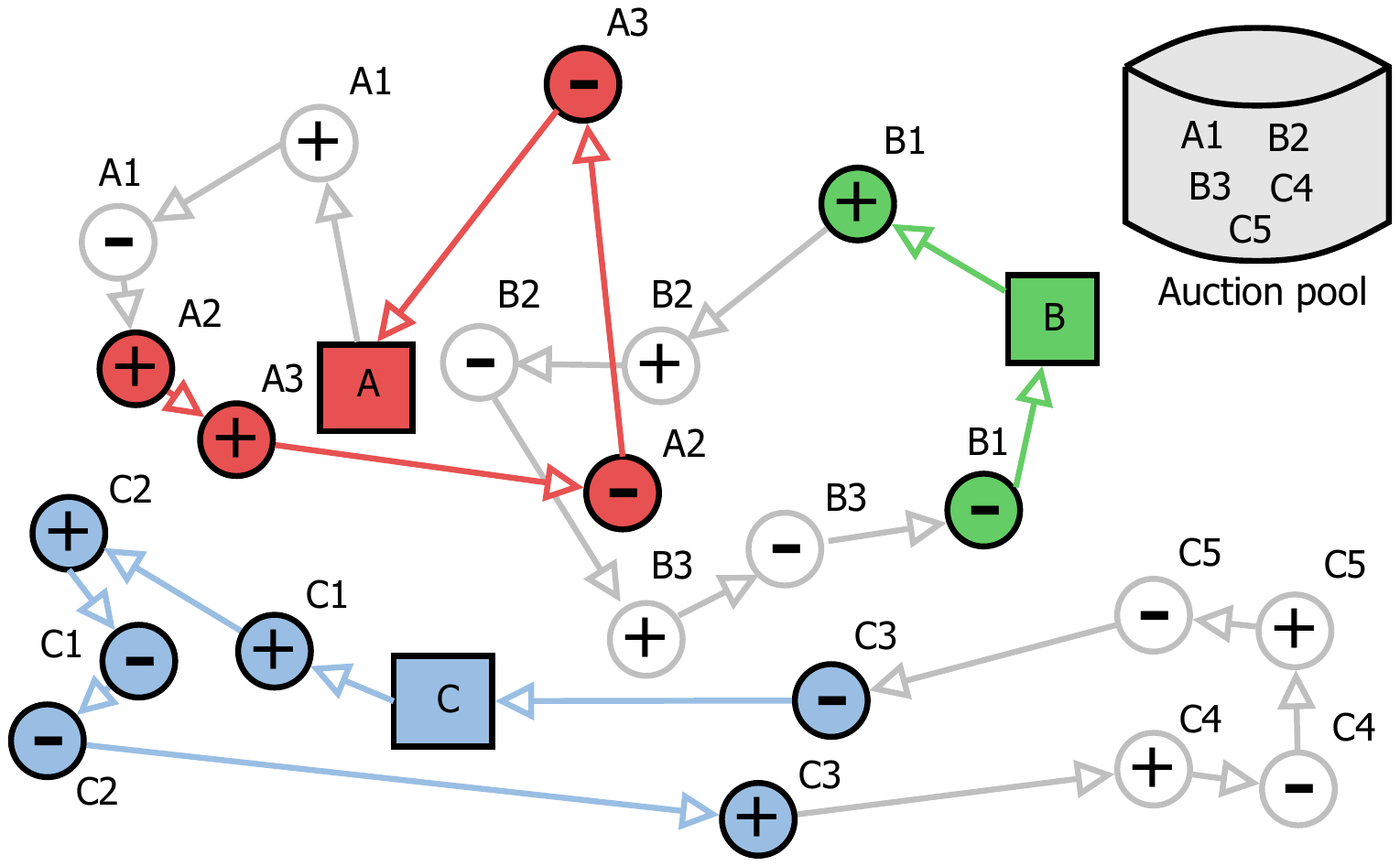}%}
\caption{Carriers submit request to the pool. Carrier A selects requests based on marginal profits, while carriers B and C take geographical information into account (\citealp{GanstererIfors}).} \label{fig:select}
\end{center}}
\end{figure}

In the first phase, participating collaborators can decide either on self-fulfillment, i.e. they plan and execute their transportation requests with their own capacities, or to offer some of them to other carriers. Aiming at network profit maximization, carriers should try to offer requests that are valuable for other network participants. Otherwise, the auction mechanism will not yield improved solutions. However, the identification of requests that are valuable for collaborators is not trivial since the actors do not want to reveal sensitive information. Different selection decisions are illustrated in Figure~\ref{fig:select}, where carrier A selects traded requests based on his own preferences, while the other carriers offer requests that have a higher probability to be attractive for their collaborators. In such a system, there is of course a high risk of strategic behavior, since carriers might increase individual benefits by offering unprofitable requests, and achieving very attractive ones in return. 

The first auction phase is investigated by \cite{Gansterer2016}. The authors show that the best request evaluation criteria take geographical aspects into account. They clearly dominate pure profit-based strategies. \cite{Schopka2017} investigate pre-selection strategies, which they classify as (i) request potential or (ii) tour potential valuation strategies.
\cite{Li2016} assume that carriers in combinatorial auctions have to solve a PDPTW with reserved requests for deciding which requests should be submitted to the auction pool.

In the second phase, the requests in the pool are grouped into bundles. These are then offered to participating carriers. If it is assumed that carriers can get a set of requests that exceeds their capacities (outsourcing option), the generation of bundles can be moved to the carriers themselves. The auctioneer could then offer the set of requests without grouping them to bundles, while the carriers give their bids on self-created packages of requests. The obvious drawback of this approach is that the auctioneer cannot guarantee to find a feasible assignment of bundles to carriers. This the reason why an outsourcing option has to be included.

A simple method to overcome strict capacities (no outsourcing), is to assume that the auctioneer assigns at most one bundle per carrier. Since carriers give their bids on bundles, they can easily communicate whether they are able to handle a specific bundle or not. However, from a practical point of view, offering all possible bundles is not manageable, since the number of bundles grows exponentially with the number of requests that are in the pool. An intuitive approach to reduce the auction's complexity is to limit the number of requests that are traded. \cite{LiRong2015} do not allow the carriers to submit more than one request to the auction pool. \cite{Xu16} show effective auction mechanisms for the truckload carrier collaboration problem with bilateral lane exchange. Again carriers offer only one lane, which is the one with the highest marginal cost. In the multi-agent framework presented by \cite{Dai2011}, there is only one request traded per auction round. Thus, it is a non-combinatorial auction, where carriers act as auctioneers when they want to outsource a request to other carriers, whereas they act as bidders when they want to acquire a request from other carriers. However, limiting the number of offered items, obviously decreases the probability to find good solutions. \cite{Gansterer2017} show that, without a loss in solution quality, the set of offered bundles can be efficiently reduced to a relatively small subset of attractive ones. They develop a proxy function for assessing the attractiveness of bundles under incomplete information. This proxy is then used in a genetic algorithms-based framework that aims at producing attractive and feasible bundles. With only a little loss in solution quality, instances can be solved in a fraction of the computational time compared to the situation where all possible bundles are evaluated. 

Complexity can also be reduced by performing multi-round auctions. These are generally intended to offer subsets of the traded items in multiple rounds. Previously gained information can be used to compose the setting for the next round. By this, the bidders are never faced with the full complexity of the auction pool.

A multi-round price-setting based combinatorial auction approach is proposed by \cite{Dai2014}. In each round of the auction, the auctioneer updates the price for serving each request based on Lagrangian relaxation. Each carrier determines its requests to be outsourced and the requests to be acquired from other carriers by solving a request selection problem based on the prices.

Regarding the last auction phase, i.e. profit sharing, we refer to a broad survey presented in \cite{Guajardosurvey}.

\cite{Chen2016} propose an alternative to combinatorial auctions for carrier collaboration, which is \textit{combinatorial clock-proxy exchange}. This exchange has two phases. The clock phase is an iterative exchange based on Lagrangian relaxation. In the proxy phase, the bids that each carrier submits are determined based on the information observed in the clock phase. 

\paragraph{Future research directions}
Auctions can be powerful mechanisms for increasing collaboration profits. However, each of the 5 auction phases bears a complex and at least partly unsolved decision problem in itself. To make auctions efficiently applicable to real-world settings, many challenging questions still have to be answered. For instance, the strong relationship between the five auction-phases (\citealp{Berger2010}) has not been investigated so far. The majority of studies focuses on one of the five decision phases, while an integrated and practically usable framework is still missing. 
Also the realistic aspect that carriers might behave strategically, opens many interesting research questions. In particular, the influence of strategic behavior in the request selection or in the bidding phase are yet to be investigated. At this point, effective profit sharing mechanisms are needed, since these have the potential to impede strategic behavior. 

So far, only relatively simple auction procedures have been investigated. It might be worth to adopt more complex mechanisms like, e.g., multi-round value-setting auctions (\citealp{Dai2014}). Also, in the literature there is no structured assessment of the potential of auctions in comparison to optimal solutions. These can of course only be guaranteed, if the auctioneer gets full information. Thus, it is worthwhile to investigate the value of information, i.e. the assessment of types or levels of information that increase solution quality.  

\section{Conclusion}
\label{concl}
Collaborative vehicle routing is an active research area of high practical importance. In this review paper, we have given a structured overview and classification of the related literature. We identified three major streams of research, which are (i) centralized planning, (ii) non-auction-based decentralized planning, and (iii) auction-based decentralized planning. Literature was further classified based on the underlying planning problem and the collaboration setting. 

We discussed recent developments and proposed future work directions, which, for instance, are
\begin{itemize}
\item the application of collaborative frameworks to more complex, e.g. multi-modal, transportation systems,
\item the investigation of strategic behavior, and effective profit sharing mechanisms to avoid it,
\item a comparative study assessing the advantages of auction-based compared to non-auction-based systems,
\item the assessment of the value of information in decentralized exchange mechanisms.
\end{itemize}

\textcolor[rgb]{0,0,0}{In order to produce comparable results, an open access repository of related benchmark instances would be helpful. This would enable a structured investigation of performance gaps between centralized and decentralized approaches. Publicly available data instances are provided by (i) \cite{Fernandez16} (\url{http://or-brescia.unibs.it/instances}), (ii) \cite{WangKG2014}, \cite{WangKopf2015} (\url{http://www.logistik.uni-bremen.de/}, and (iii) \cite{Gansterer2017} (\url{https://tinyurl.com/y85hcpry}).}

\section*{Acknowledgements}%
This work is supported by FWF the Austrian Science Fund (Projectnumber P27858-G27)



\section*{References}
  \bibliographystyle{elsarticle-harv} 
 
	\bibliography{CollVRP}

\end{document}